\documentclass[compsoc,10pt]{IEEEtran} 

\usepackage{graphicx, epstopdf,  cite}

\graphicspath{{img/}}

\begin{document}

\title{The Interplay of Reconfigurable Intelligent Surfaces and Mobile Edge Computing in Future Wireless Networks: A Win-Win Strategy to 6G}

\author{Mithun Mukherjee,~\IEEEmembership{Senior Member,~IEEE,} Vikas Kumar,~\IEEEmembership{Member,~IEEE,} \\Mian Guo,~\IEEEmembership{Member,~IEEE,} Daniel Benevides da Costa, \IEEEmembership{Senior Member, IEEE}, \\Ertugrul Basar, \IEEEmembership{Senior Member, IEEE, and Zhiguo Ding, \IEEEmembership{Fellow, IEEE}}

\thanks{M. Mukherjee is with the School of Artificial Intelligence, Nanjing University of Information Science and Technology, Nanjing 210044, China (e-mail: m.mukherjee@ieee.org).}
\thanks{V. Kumar is with the Bharat Sanchar Nigam Limited, India (e-mail: vikas.kr@bsnl.co.in).}
\thanks{M.~Guo is with the Guangdong Polytechnic Normal University, China (e-mail: mian.guo@ieee.org)}
\thanks{D. B. da Costa is with the Future Technology Research Center, National Yunlin University of Science and Technology, Douliu, Yunlin 64002, Taiwan, R.O.C. and with the Department of Computer Engineering, Federal University of Cear\'{a}, Sobral 62010-560, CE, Brazil (e-mail: danielbcosta@ieee.org).
}
\thanks{E. Basar is with the Electrical and Electronics Engineering, Koc University, Istanbul, Turkey
(e-mail: ebasar@ku.edu.tr).}
\thanks{Z. Ding is with the Department of Electrical and Electronic Engineering, The University of Manchester, UK  (e-mail: zhiguo.ding@manchester.ac.uk).}
}
\maketitle

\begin{abstract}
Reconfigurable intelligent surface (RIS)-empowered communication is being considered as an enabling technology for sixth generation (6G) wireless networks. The key idea of RIS-assisted communication is to enhance the capacity,  coverage,  energy efficiency, physical layer security, and many other aspects of modern wireless networks. At the same time, mobile edge computing (MEC) has already shown its huge potential by extending the computation, communication, and caching capabilities of a standalone cloud server to the network edge. In this article, we first provide an overview of how MEC and RIS can benefit each other. We envision that the integration of MEC and RIS will bring an unprecedented transformation to the future evolution of wireless networks. We provide a system-level perspective on the MEC-aided RIS (and RIS-assisted MEC) that will evolve wireless network towards 6G. We also outline some of the fundamental challenges that pertain to the implementation of MEC-aided RIS (and RIS-assisted MEC) networks. Finally, the key research trends in the RIS-assisted MEC are discussed.
\end{abstract}

\section{Introduction}
While fifth generation (5G) technologies have started to roll out over the world, the buzz words 5G++, beyond 5G, and more specifically sixth generation (6G) are knocking at the door. Researchers and experts from academia and industry are in pursuit of figuring out what 6G will be, as well as the new and advanced technologies, system architectures, and use-cases. Several leading companies and top research institutes (such as 6Genesis of Finland, TOWS for 6G LiFi in the UK, 5G Lab Germany in Dresden, 6G Mobile Communications R\&I Lab at the University of Sussex, and 5G innovation facility in University of Technology Sydney, Australia)  have set up 6G research centers for rapid development towards this paradigm shift with a holistic approach.

\subsection{Mobile Edge Computing: Current Trends}
To catch up with this trend towards 6G, in recent years, several industries are focusing their technological advancement towards the high performance in cloud data centers. For example, in 2020, NVDIA announced the potential use of DGX A100 NVIDIA’s third generation Artificial Intelligence (AI) system box~\cite{NVIDIADGX} that is aimed for massive gain in performance for AI-related and cutting-edge applications with  low power consumption. At the same time, we are witnessing a change in paradigm that constitutes a well-run centralized data center infrastructure to the network edge, particularly, when there is a need to deliver proximity, low-latency, and reliable services for the mission-critical applications, such as remote-surgery, industrial automation and driverless cars. The leading industries with their cloud service providers (e.g., EGX Edge AI platform, NVIDIA RTX graphics with CloudXR, GPU virtualization, and Qualcomm Technologies’ Boundless XR client optimizations~\cite{QualcommXR}, and EdgeConneX~\cite{EdgeConnecX}) are making their way for the deployment of edge-assisted service provisioning. 

The basic concept of edge computing is to bring the functional capabilities of cloud computing close to the source of the data generations, i.e., to the network edge. By extending the computational and the storage resources to the resource-constrained end-users, edge computing greatly removes the bottleneck of the transmission delay between the end-users and the remote cloud data centers. However, at the same time, the edge server inherently suffers from the following issues. A standalone edge server has limited computational and storage resources compared to the resource-rich cloud server. Therefore, apart from the computational resource allocation, the placement of task processing (i.e., process at its own server, push to the cloud, or migrate to another edge server) needs communication resources. Moreover, one cannot ignore the demand for wireless communication resources when the end-users ask for edge computing.

\begin{figure}[t]
\centering
\includegraphics[width=0.47\textwidth]{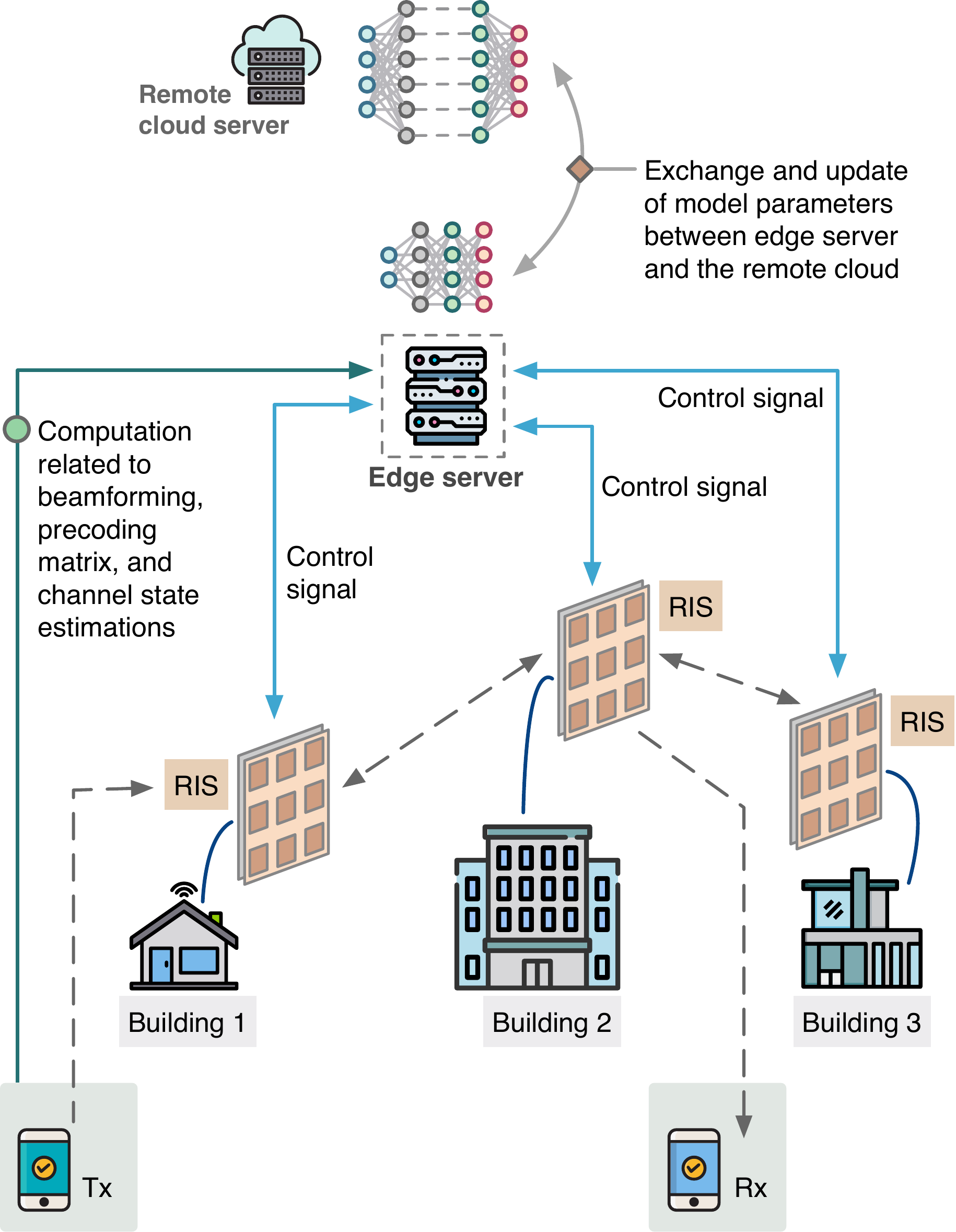}
\caption{Potential advantages of edge computing in a network with multiple RISs: i) Computing, caching, and storage  capabilities of MEC can assist the RIS controller to reconfigure the phase shift of the RIS elements and ii) The computation related to the beamforming, precoding, and channel state acquisition can be offloaded to the edge server.}
\label{Fig:MECforIRS}
\end{figure}

\subsection{Reconfigurable Intelligent Surfaces (RISs): Fundamentals}
The advancement in programmable meta-materials has sparked an emerging technology, reconfigurable intelligent surfaces (RISs)-assisted communication~\cite{BasarRIS2019}, as a part of beyond-5G and towards 6G. With other terminologies, such as intelligent reflecting surfaces (IRSs), passive large intelligent surfaces (LISs), and software-controlled metasurfaces, the main idea of RISs is to manipulate the scattering properties of electromagnetic waves through the use of passive \emph{elements} to enhance the \emph{quality} of wireless communications. To this end, several advanced signal processing techniques, such as interference nulling (and cancellation), creation of virtual line-of-sight links, and new forms of physical layer security can be implemented. Compared with the point-to-point communication, in RIS-assisted wireless communication systems, each element of the RIS receives the incoming signal, in a superimposed manner, from the transmitter, and then it scatters or reflects back this signal to the destination while adjusting the amplitude, frequency, delay, and polarization of incoming electromagnetic waves. These RIS-assisted communication systems have promising potential applications in multi-user systems and non-orthogonal multiple access with hybrid active and passive beamforming. Interestingly, RIS-based modulation and data transfer are receiving a great deal of attention in the wireless research community.

The role of RISs has been recently studied in MEC systems (cf.~\cite{bai2021empowering} and references therein), where a couple of use-cases are identified. At the time of writing the article, several 6G initiatives and white papers discuss the speculative and yet to be explored technologies.  In this context, we aim to present an overview of how these two enabling technologies, MEC and RIS, can emerge in harmony as a potential candidate in 6G. Specifically, we discuss the benefits, challenges, and potential use-cases in MEC-aided RIS (and RIS-assisted MEC) networks towards 6G, and present open research problems.

\section{An interplay between RIS and MEC}
This section presents how MEC can benefit RISs and how RISs can assist MEC in overcoming the inherent problems towards data-intensive, low-latency, high-reliable, and secure 6G wireless networks.

\begin{figure*}[t]
\centering
\includegraphics[width=.96\textwidth]{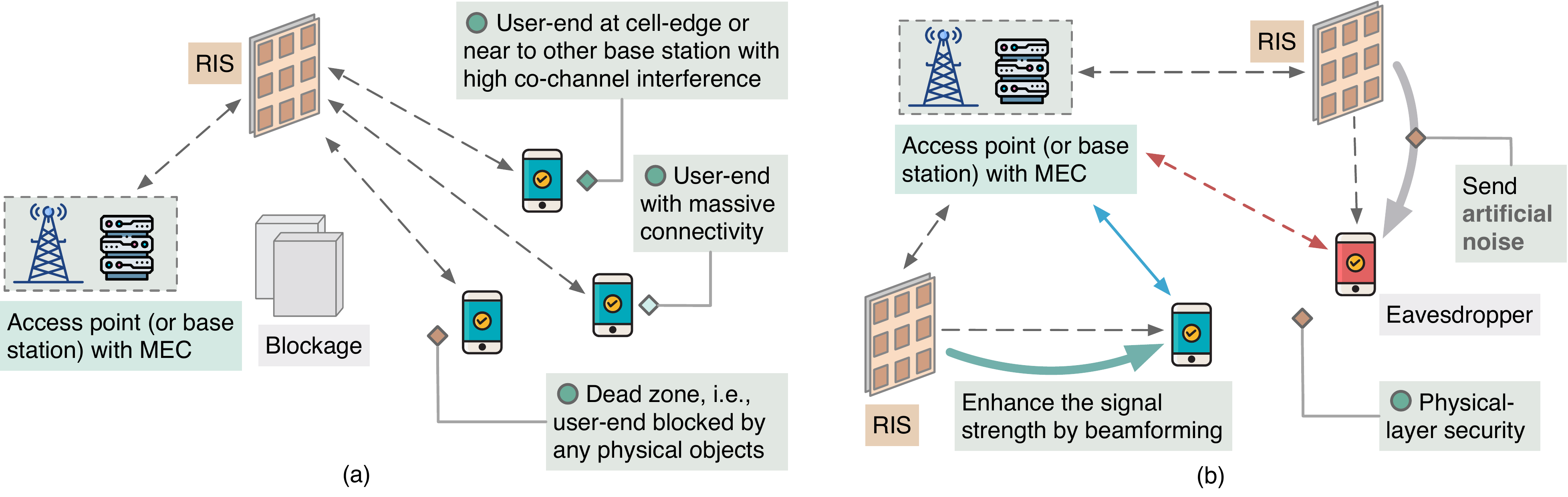}
\caption{Some use-cases of RIS-assisted MEC. (a)  Near to other interference zone coverage, massive connectivity, and dead-zone coverage, and (b) Secure physical communications.}
\label{Fig:RISforMEC}
\end{figure*}

\subsection{MEC-Assisted RIS}
\subsubsection{Computation Related to Complicated Optimization Problems in RIS-enabled Networks: Can MEC Assist?}

In practice, an RIS can be installed or placed in many different locations, such as surfaces of the indoor ceiling and outdoor building,  lamp posts, road signs, and movable objects (e.g., unmanned aerial vehicles and air balloons). This is indeed a practical way to establish a reflected path with beamforming between the source and destination. To illustrate this scenario, Fig.~1 depicts multiple RISs installed in different building surfaces along with a transmitter-receiver pair. These RISs collaboratively assist the signal transmission between the transmitter and the receiver through their intelligent reflections. Now, \emph{how to optimize the beamforming vector and RIS scattering coefficients} is a challenging factor. It becomes more critical when these optimizations demand real-time output. These optimizations often require huge computation due to the massive end-user connectivity and mobility of the end-users and time-varying channel conditions. In addition, the number of elements in RIS and other properties such as the RIS locations, connectivity, and computation and communications resources of the access points/base station and  user-ends (UEs) make the optimization problem more complex. 

Although a centralized remote cloud server can easily handle huge computation, an edge server inherently provides the advantage of distributed computing closer to the network edge. Therefore, it is not possible to always rely on cloud servers that are physically apart and suffer from significant fronthaul/backhaul delay. In this way, the key solution is to exploit the \emph{computing resources of the MEC to reconfigure the RIS whenever necessary}. As shown in Fig.~\ref{Fig:MECforIRS}, one can place an edge server to handle these complicated computations related to beamformimg, precoding, and channel state acquisition for the collaborative transmission with RISs. This edge server basically assists the RIS controller in generating the control signal. When needed, the edge server also obtains high computing resources of the remote cloud server while exchanging the parameters for computation purposes.

\subsubsection{Challenges in AI/ML-enabled RIS: Motivation to Use MEC in RIS}
Over the past decade, artificial intelligence/machine learning (AI/ML) is being widely investigated in various kinds of system design and optimization tasks for future wireless communication systems. Following the trend, AI/ML can also be leveraged in RIS-enabled networks. In the existing literature, active and passive RISs are studied. Now, on the one hand, when active elements are considered, how to reconfigure the behavior of the elements, i.e., the cross-connection between the elements, optimal phase angles of the RIS elements as well as minimizing the power consumption are the major design issues. On the other hand, when the passive elements are used mainly due to their advantage of cost-saving and easy deployment, the channel estimation at the receiver and the RIS end, the beamforming vector, and the precoding matrix generation at the transmitter are essential. Irrespective of the element type, one cannot ignore the fact that the optimization problems become non-trivial due to extremely high-dimensional optimization. Therefore, to address the above challenge, AI/ML is a real-time efficient tool for being employed in the RIS-enabled networks.

The offline approaches used for channel estimation, beamforming matrix formation, and reconfiguration of RIS elements involve a significant amount of computation for training purposes. In addition, the reinforcement learning-based approaches also require a computation unit. Therefore, there is an obvious need for computing units to leverage AI/ML in a RIS-enabled network fully. In this context, MEC can help to bring the computational, storage, and caching capabilities towards the network edge. The computing units that run the AI/ML model can be benefited from the edge computing concept. Basically, the system-level implementation of AI/ML in RIS-enabled networks, the computing, and storage power of edge servers can be fully utilized. As illustrated in Fig.~\ref{Fig:MECforIRS}, the remote cloud server can act as massive computing and storage-rich place, thus when required, the exchange and update of model parameters of the AI/ML can be performed between the edge server and the remote cloud server. 

\subsection{RIS-assisted MEC}
\subsubsection{Improvement in Wireless Link Quality}
Although MEC brings computational, caching, and storage resources towards the network edge, the connectivity, and coverage of the access points/base station play a critical role. To say, when a UE aims to avail the computing, caching, and storage resources of edge server and or download the results. Depending on the application type, the uploading and downloading size vary. However, irrespective of application type and service, the uploading/downloading time plays a major role in delay-sensitive service provisioning. This becomes worse when the network coverage is in poor locations near the cell-edge and blocked by obstacles. The uploading and downloading success rate and the latency are significantly affected by the communication resource allocation. This, in turn, affects the \emph{computation resource} allocation in terms of either CPU frequency allocation or computation time allocation of the edge server. The main reason is that transmission delay (i.e., uploading and downloading time) and task execution time at the edge server have joint impact on the delay and reliability in MEC-enabled networks. To address the above shortcoming that arises due to the connectivity and coverage of the network services, RIS can assist MEC. Fig.~\ref{Fig:RISforMEC}(a) depicts a scenario where RISs can assist MEC in several situations, such as massive connectivity, dead-zone coverage, and near to other interference zone coverage.

\subsubsection{Energy Efficient MEC}
Energy consumption stemming from the upload and download of the task data is always a challenging factor when UE aims to offload the task data to the edge server. With the assistance of RISs, due to properly designed beamforming vectors, the transmit power consumption can be lowered at the UE and edge server’s side while offloading, obviously with the constraints such as computing capacity of the edge server, timeliness, and reliability requirements of the applications. Undoubtedly, wireless power transfer assisted by RISs~\cite{Bai_Nallanathan_TWC2021} will open up a new domain to improve the energy efficiency in edge computing for future networks.

\subsubsection{Secure Edge Computing}
\begin{figure}[t]
\centering
\includegraphics[width=0.485\textwidth]{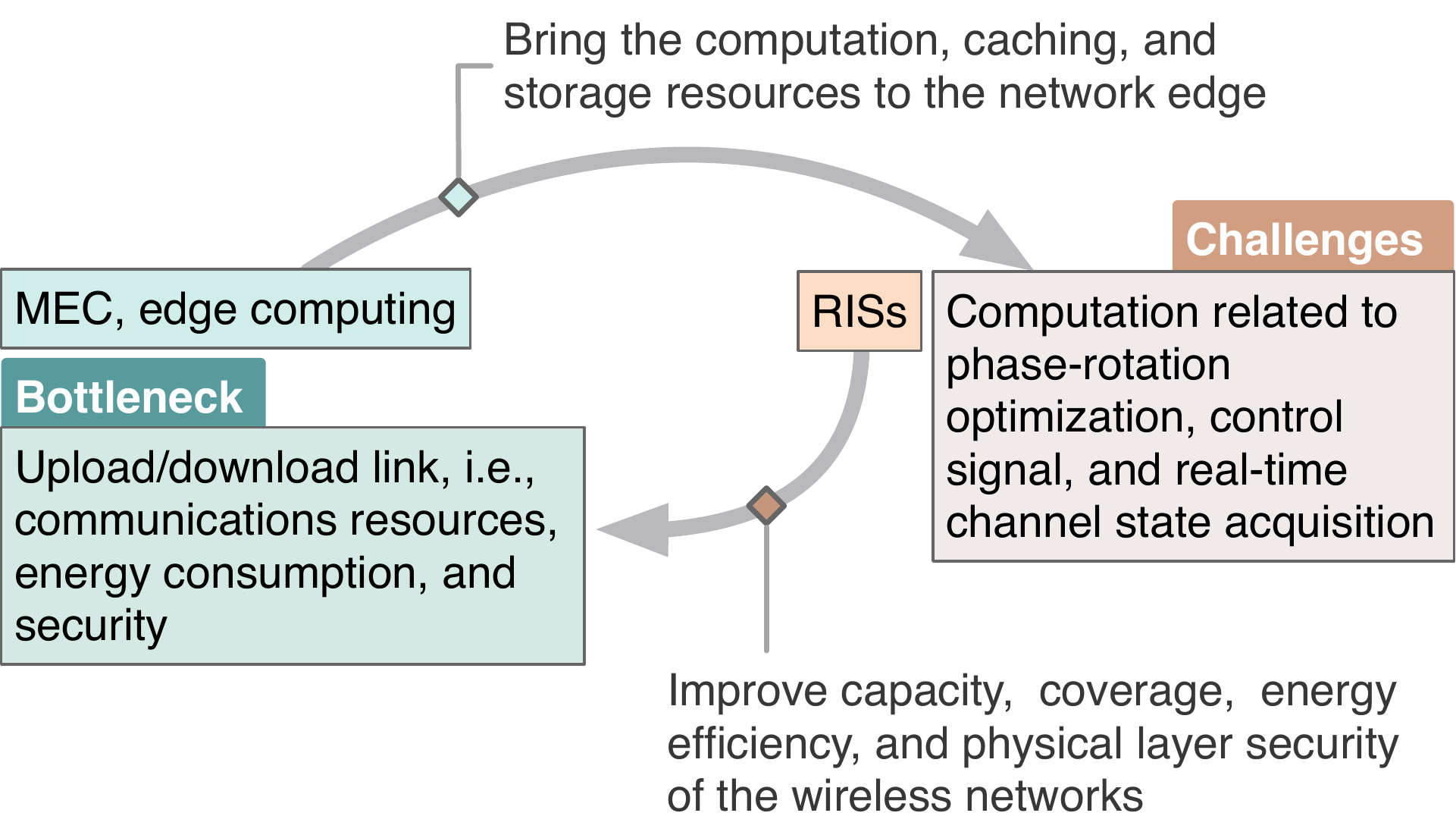}
\caption{An illustration on how MEC and RIS can benefit each other.}
\label{Fig3}
\end{figure}

Physical layer security has already shown its importance in 5G due to its inherent properties, such as no secret keys and less complex cryptographic algorithm, by exploiting the advantages of the dynamic nature of the wireless environment. It becomes easier to cancel out the received signal strength for the eavesdroppers via placing RIS near them by exploiting the statistical channel state information. It is interesting to observe that with the assistance of RIS, an artificial noise~\cite{RuiZhangWCLIRSSecurity} can be introduced to suppress the signal strength of the eavesdroppers, thereby improving the secrecy performance of the wireless networks (see Fig.~\ref{Fig:RISforMEC}(b)). In addition, it is also possible to establish \emph{secure} communication with legitimate users by properly designing beamforming matrix while increasing the signal strength for the legitimate users. Recent studies demonstrate how RISs can be leveraged for covert communication~\cite{DusitCovert2020} to hide the existence of physical-layer signal transmission from eavesdroppers. Using this kind of new setup with RIS, the physical layer security has huge potential benefits in edge computing. For example, the legitimate users can be protected from eavesdropping while using the services provided by the edge server. Moreover, this physical layer security will bring an additional dimension to the secure edge computing, particularly in data transfer among edge servers, cloud servers, and UEs. To summarize, in Fig.~\ref{Fig3}, we show the beneficial role of RIS and MEC that opens up future research opportunities.

\section{Potential MEC/RIS Use-case Scenarios for Future Wireless Networks}
In this section, we discuss the use-cases of MEC/RIS systems for next generation wireless networks. 
\begin{figure*}[t]
\centering
\includegraphics[width=.99\textwidth]{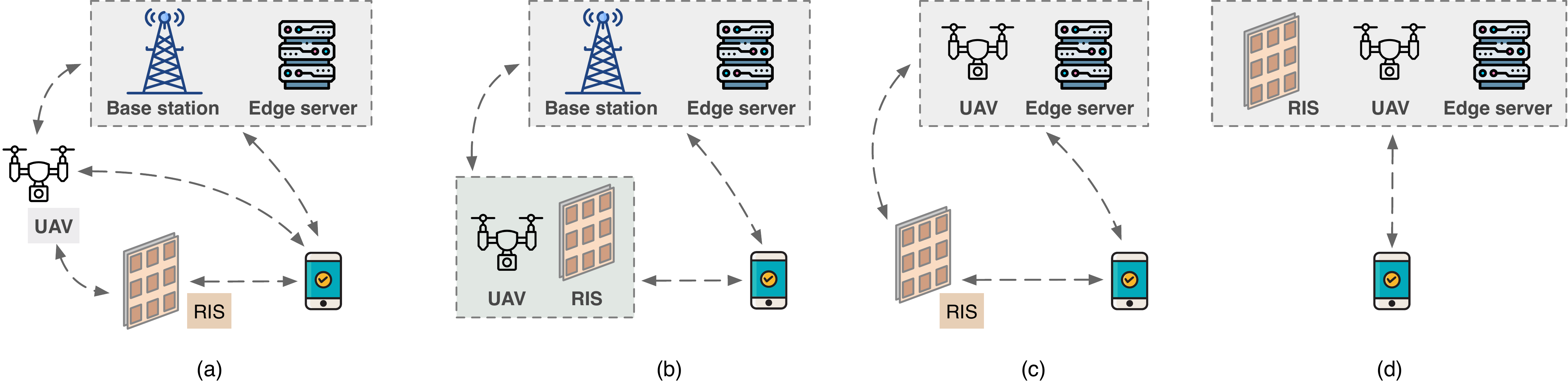}
\caption{Potential MEC/RISs use-case scenarios with flying units for future wireless networks. (a) RISs installed on the building surface act as assistive units between UAV and UE, (b) RISs are installed in the UAV, (c) RISs installed on the building surface support the communication between MEC-assisted UAV, and (d) RISs are placed in MEC-assisted UAV.}
\label{Fig4:UAVandAVR}
\end{figure*}

\subsection{MEC/RIS-assisted Flying Units}
At the final stage of 5G deployment, unmanned aerial vehicles (UAVs), as flying units, are being considered as one of the potential solutions to the scenarios where the coverage and connectivity are limited. These UAVs can reside between the MEC-enabled base station/access points and UEs. In particular, UAVs aim to enhance overall network performance in terms of energy efficiency and the high data rate, particularly when the UEs are distant from the base station/access points~\cite{MarcoUAV-TVT:2020}. However, sometimes the connectivity from the UAV to the ground UEs is blocked by the obstacles. In this case, an RIS installed on the building surface can assist the UAV in overcoming this issue. It is also possible that these RISs could be installed in the UAV rather than on the building surface. Apart from the fact that a UAV can act as a flying base station, it can be equipped with a computing unit to provide additional computing, caching, and storage resources to the UEs. In this setup, an RIS can assist the edge computing-supported UAV either as a separate unit installed on the building surface or integrated with the UAV itself. Fig.~\ref{Fig4:UAVandAVR} illustrates some of these configurations with flying units, RISs, and edge server. These setups will be quite disruptive and have their benefits for realizing 6G technologies in the future. 

\subsection{RIS-assisted MEC to Support Augmented Virtual Reality in 6G}
A huge expectation has been put forward to see the next form of augmented virtual reality (AVR) in 6G. Some of the potential applications are remote surgery, holographic projection, networked games, remote automation, connected and autonomous car, and tactile/haptic communications. To meet diverse requirements of these exciting technologies with a varied degree of connectivity, latency, and reliability, edge computing plays a major role. For example, we take the following edge computing system in~\cite{Hua_Green2021} where a UE has inference tasks (i.e., image processing) to be processed as a part of AI/ML-related service. To execute, the UE first has to upload its raw data to the edge server, and then the edge server obtains the inference results by feeding the data to an already trained machine-learning model. However, the delay due to uploading and downloading raw data and model parameters limits the performance. To elevate this, an RIS can assist uplink and downlink transmission.
 
Recently, over-the-air computation (AirComp) that integrates communication and computation attracts academia and industries' attention due to its fast data aggregation from IoT devices. However, due to the unreliable channel conditions, the performance of AirComp is severely limited. To address this, an RIS has been exploited in~\cite{IRSWework13}. Interestingly, the authors suggested a data-driven optimization problem without any prior knowledge of channel distribution. Therefore, it is evident that RIS-assisted MEC has a full potential to realize several emerging and yet to be discovered use-cases in 6G.

\subsection{Coexistence with Other Wireless Technologies}
In 6G, it is evident that several technologies will co-exist over the networks. Some of them are wireless power transfer, cell-free network, and power-, code-domain, and time-sharing non-orthogonal multiple access.  How to model this hybrid network while taking full advantage of their characteristics combined with RISs and MEC will be a priority in 6G. A recent study in~\cite{Zhou_RuiZhang_WCL2021} suggested a flexible time-sharing NOMA scheme for MEC systems by jointly designing RIS phase shifts and other related parameters of the NOMA-based system to minimize the latency which is basically calculated as the sum of two UE's computation offloading time. Several leading research groups and industrial experts are advocating towards using sub-millimeter, often termed as \emph{whisper radio}, and terahertz frequency band in 6G for ultra-speed and low-latency communications in 6G. However, such frequencies are susceptible to high propagation attenuation and blockage.  An interesting and detailed study on how RISs can directly affect the task offloading chances for the UEs that suffer from mm-Wave link blockage has been presented in~\cite{Cao_mmWave_2021}. These recent studies indicate an encouraging result, that is, an RIS is capable of assisting mmWave and THz communication in edge computing systems. Subsequently, many open research challenges, such as how to design beamforming with less complexity, how to avoid the obstacles, are yet to be investigated.

\section{Illustrative Simulation Results}
We take an MEC scenario with multiple UEs and one edge server. In this setup, an RIS is employed to improve the average delay performance of the network. Due to the limited local computing resources and the large input data size, these UEs aim to offload their task data to the edge server. We assume that 2/3 of the UEs are in good network coverage area, while the rest of the UEs suffer from poor network coverage. It is also considered that computing resource in edge server is $7\times10^9$ cycles/slot, UE's local CPU speed is uniformly distributed over $[100, 1000] $ MHz, the input data of the end-users device is uniformly distributed over $[0.2, 0.7]$ Mbits, and the number of CPU cycles per bit is 500 cycles/bit. We set the total available bandwidth as $1$ MHz, which is equally distributed among all UEs. The received signal power from the UEs subject to good and poor network coverage is assumed to be uniformly distributed over $[-50, -30]$ dBm and $[-110, -95]$ dBm, respectively. Further, taking the path loss models in~\cite{Tang2020}, the improved received signal power by large and small RIS-assisted beamforming is uniformly distributed over $[-90, -65]$ dBm and $[-65, -50]$ dBm, respectively. We further assume that the final output data size of these tasks is very low compared to the input data size, therefore, the downloading time for the output from the edge server to the UEs is ignored. 
\begin{figure}[t]
\centering
{\includegraphics[width=0.48\textwidth]{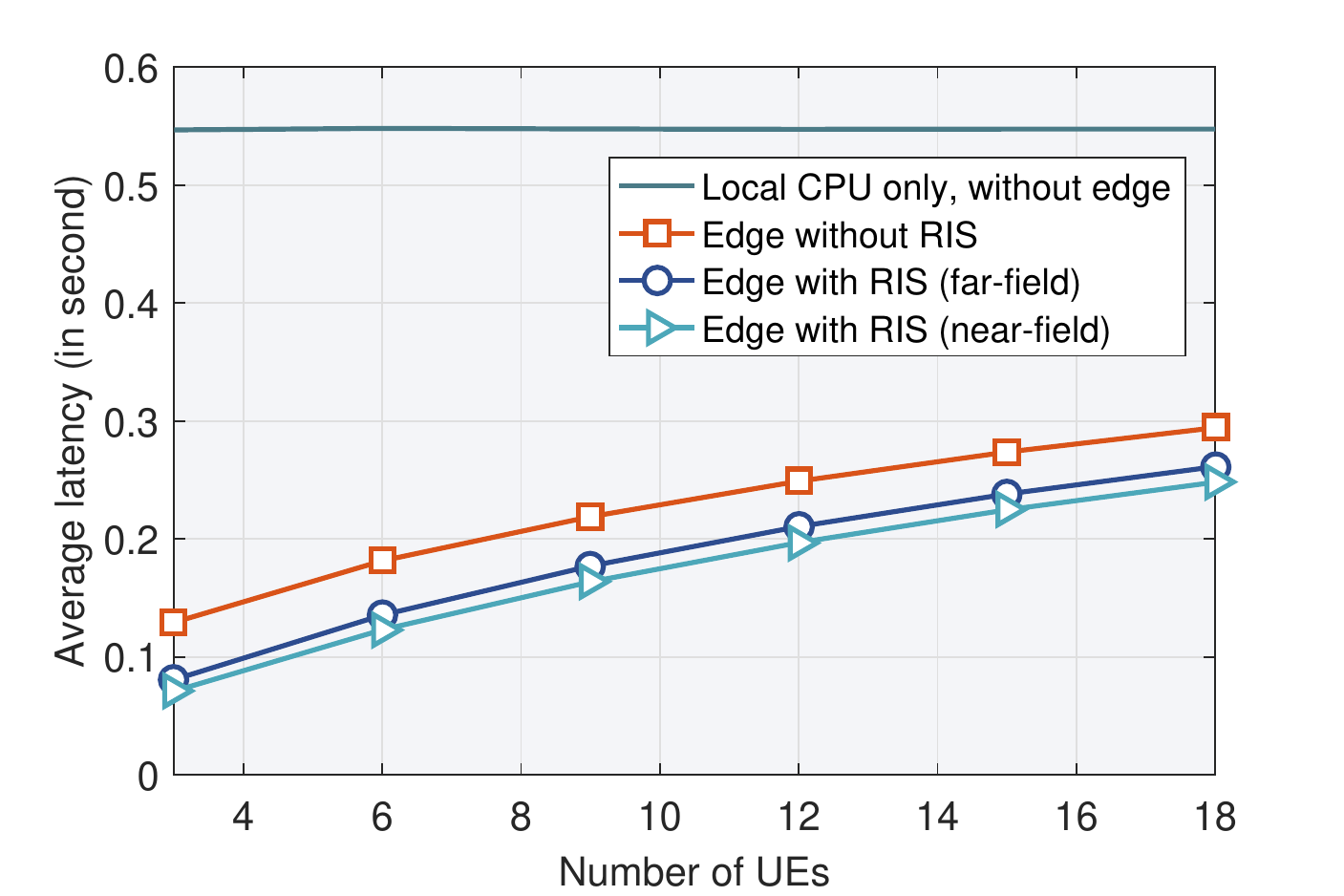}
\label{fig:AvgDelayBeamformingUser}}
\caption{Average latency versus the number of UEs.} 
\label{fig:Broadcasting}
\vspace{-10pt}
\end{figure}
\begin{figure*}[ht]
\centering
\includegraphics[width=\textwidth]{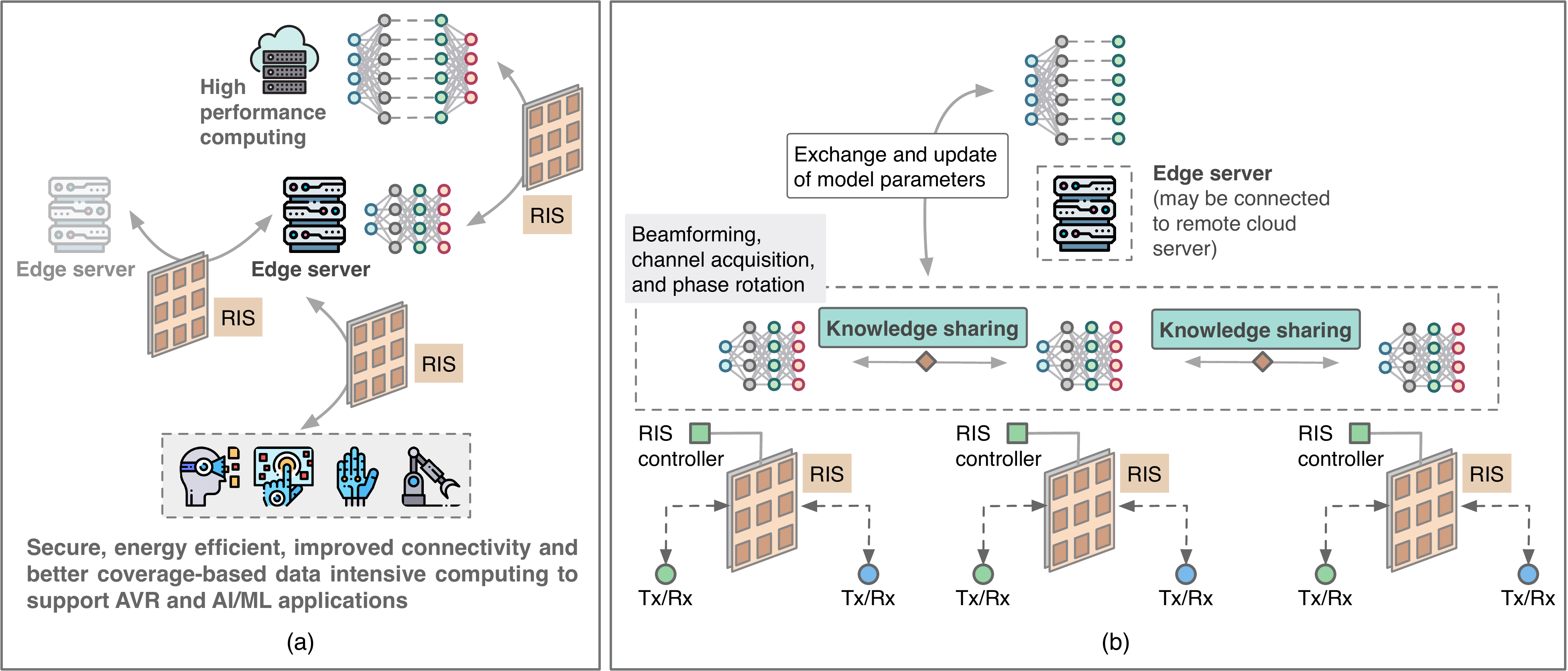}
\caption{(a) RISs assist the communication resource allocation to support secure and energy-efficient MEC-enabled AVR and AI/ML applications. (b) Knowledge sharing in RIS/MEC networks with federated learning. The edge server exchanges and updates a global model for RIS configuration, while AI/ML-enabled RIS controllers train local models with local data. The global and local models are shared and iteratively updated until the global consensus model is achieved. Then, all RIS controllers can use the shared model/knowledge.}
\label{Fig6:KnowledgeSharing}
\end{figure*}
Moreover, the noise power spectrum density at the edge server is set as $-174$ dBm$/$Hz. Fig.~\ref{fig:Broadcasting} illustrates the average delay performance over the network. We observe that the average delay increases with the increase of the UEs. The main reason is as follows. With the increase of the number of UEs, the transmission delay to send the data to the edge server increases mainly due to the fact that the uploading bandwidth is equally shared by the UEs. Therefore, the average delay increases with the increase of transmission delay. Note that the performance of the UEs under poor coverage area has a strong influence on the average delay performance of the network. Subsequently, improving the uploading transmission rate between the UEs in poor network coverage area and the edge server with RIS assistance results in reduced transmission delay. Therefore, the average delay decreases. 
Two different setups with RIS are considered: one refers to an edge server with far-RIS and another is an edge server assisted by near-RIS. It is used far-RIS to assist long-distance communication between UE and edge server while the near-RIS boosts short-distance communication between UE and edge server. Basically, the received signal power at the edge server with far-RIS is lower than the edge 
server with near-RIS due to the far-field and near-field propagation features, being in accordance with the insights attained in [14]. Therefore, the achievable data rate for uploading under far-RIS regime is lower than the near-RIS case. As a result, the average latency performance with far-RIS is slightly lower than near-RIS-assisted edge server. Interestingly, from the figure, one can see that both far- and near-RIS-assisted systems outperform the system without RISs.

\section{Research Challenges and Future Directions}
In this section, we outline the future research directions for designing RIS-assisted MEC (and MEC-aided RIS) systems in future wireless communication.

\subsection{Reflection Coefficient Design Issues, Location and Number of RISs}
The location of RISs~\cite{RISPosition2018} and the number of elements in each RIS are being considered as main design criteria for any RIS-assisted network. Irrespective of metamaterial or patch-array in RIS, it will always be challenging to address the hardware-related issues, phase-shift, and reflection co-efficient optimization. Subsequently, it is well understood that if the whole RIS-assisted network area is further divided into sub-zones, the effectiveness of using RISs in MEC will be further increased. However, this performance improvement comes with the expense of hardware complexity and computational burden due to beamforming and other related factors. In addition, the material cost and the energy consumption issue due to the deployment of RIS would be kept in mind. The computation resource allocation in MEC solely depends on how we allocate the communication resources in RIS-assisted networks.  Therefore, the optimizations of reflection coefficients, number, and location of these RISs and other network elements (base station/access point, edge servers, and UEs) will be challenging, if not impossible.  

\subsection{Knowledge-Sharing in RIS/MEC}
There is currently compelling need of distributed AI/ML algorithms that can be deployed and  optimized using small amount of data and that can quickly converge in a time shorter than the coherence time of the wireless environment. In recent years, federated learning as a distributive learning paradigm  has been widely used in MEC for  applications that are sensitive to data privacy and learning delay. Generally, distributive learning requires many communication rounds among UEs, edge servers, and high performance cloud servers, resulting communication resource allocation and energy consumption issues (see Fig.~6(a)). To address these challenges, RISs can assist the communication resource allocation and energy consumption optimization, thereby boosting the performance of distributed learning in MEC.

In addition, as discussed earlier, MEC can also assist RIS configuration. Irrespective of what types (active, passive or hybrid) of RISs are used, complicated computation for RIS configuration is hard to ignore. Again, it is obvious that RISs will perform in a collaborative manner, which makes the \emph{reconfiguration} process more complex. To this end, AI/ML will play a critical role in reconfiguring these RISs based on the channel state information, location of the network elements, and, more importantly, the requirements imposed by the applications. In this context, the distributed approaches, e.g., federated learning, are appealing for RIS configuration with a large number of network elements (RISs, transmitter and receiver pairs, access point/base stations, and edge servers), dynamic nature of the network, and demand for the real-time response from these RIS-enable setups. Fig.~\ref{Fig6:KnowledgeSharing}(b) depicts a schematic representation of how the knowledge sharing among these RIS units will be leveraged based on federated learning. Particularly, the transfer federated learning, where the data space and even feature spaces are different and the learned common representation is used, has a huge potential to be applied in RIS/MEC networks. At the same time, the challenging factor is the updating frequency of the model parameters in the AI/ML-based knowledge- and data-driven approaches in this emerging technology in 6G.

In addition, transfer learning arises as a suitable approach in order to reduce the amount of data for system optimization in RIS/MEC systems because it combines together model-based and data-driven optimization methods. The idea is to exploit prior knowledge of the system based on mathematical methods as the initialization point, from which ML methods start interacting with the environment for system optimization. The initial network status obtained from a model embeds many of the most important features of the actual system. Relying on transfer learning approaches, it will take less time and data for ML methods to converge towards the optimal operating point. Transfer learning can be successfully used to facilitate the implementation of deep learning algorithms, especially by reducing the amount of data to be acquired for training and validation purposes.

\subsection{Security and Privacy-related Issues in RIS-assisted MEC}
We have witnessed that almost every technological advancement in the past and present paradigms poses security and privacy challenges. Although it is too early to say what will be the security and privacy issues particularly for 6G, they will be obviously challenging due to the data-intensive use-cases. The physical layer security with RISs will be one of the effective approaches by suppressing the signal strength of the eavesdroppers. At the same time, the design of the beamforming vector for legitimate and eavesdroppers,  determination of other performance metrics apart from the secrecy outage probability, as well as the security and privacy measurements to be taken are yet to be investigated with hardware impairments.

\section{Conclusions}
In this article, we have provided an overview of reconfigurable intelligent surface-enabled mobile edge computing towards the future wireless network, namely 5G and beyond. Since the research on 6G is becoming a focus of attraction among academia and industries, it is a highly timely topic to figure out the enabling technologies for 6G. In this context, this article aims to provide a helpful guide to understand how RIS and MEC can benefit each other to establish a highly efficient omnipresent network. We further outline some of the leading research challenges and provide the research trends that will enable the unforeseen potential use-case and the architectural advancement towards 6G. 

\bibliographystyle{IEEEtran}
\bibliography{IRS.bib}
\end{document}